\documentclass[twocolumn,journal]{IEEEtran}

\usepackage[T1]{fontenc}
\usepackage[cmex10]{amsmath}
\usepackage{amssymb}
\usepackage{amsthm}
\usepackage{acronym}
\usepackage{threeparttable}
\usepackage{booktabs}
\usepackage[pdftex]{graphicx}
\usepackage{cite}
\usepackage{color}

\usepackage{comment}

\usepackage{subcaption}
\usepackage{float}
\usepackage[font=small]{caption}
\usepackage[unicode=true,
 bookmarks=true,bookmarksnumbered=true,bookmarksopen=true,bookmarksopenlevel=1,
 breaklinks=false,pdfborder={0 0 0},pdfborderstyle={},backref=false,colorlinks=false]
 {hyperref}
\hypersetup{pdftitle={Your Title},
 pdfauthor={Your Name},
 pdfpagelayout=OneColumn, pdfnewwindow=true, pdfstartview=XYZ, plainpages=false}

\newcommand{\ii}{\imath}
\newcommand{\cmax}{\dot y_{\max}}
\newcommand{\cmin}{\dot y_{\min}}

\acrodef{rl}[RL]{Rate Limiter}
\acrodef{der}[DER]{distributed energy resource}
\acrodef{vsi}[VSI]{voltage source inverter}
\acrodef{gfl}[GFL]{grid-following}

\makeatletter

\let\oldforeign@language\foreign@language
\DeclareRobustCommand{\foreign@language}[1]{%
  \lowercase{\oldforeign@language{#1}}}
\theoremstyle{plain}

\theoremstyle{remark}

\makeatother

\title{Smooth Rate Limiter Model for Power System Stability Analysis and Control} 

\author{Zaint A. Alexakis,~\IEEEmembership{IEEE Student Member}, Panos C. Papageorgiou,~\IEEEmembership{IEEE Member},\\
Antonio T. Alexandridis,~\IEEEmembership{IEEE Life Member}, 
Federico Milano,~\IEEEmembership{IEEE Fellow},\\
and Georgios Tzounas,~\IEEEmembership{IEEE Member}\thanks{Z.~Alexakis, P. Papageorgiou and A. Alexandridis are with the Department of Electrical and Computer Engineering, University of Patras, Patra, Greece.  G.~Tzounas and F.~Milano are with the School of Electrical and Electronic Engineering, University College Dublin, Dublin, D04V1W8, Ireland. \\
Corresponding author's e-mail: \protect\href{http://zaintalexakis@gmail.com}{z.alexakis@ece.upatras.gr}.}
\vspace{-7mm}}

\begin{document}

\maketitle
\pagestyle{plain} 
\IEEEpeerreviewmaketitle

\begin{abstract}
  The letter proposes a smooth \ac{rl} model for power system stability analysis and control.  The proposed model enables the effects of derivative bounds to be incorporated into system eigenvalue analysis, while replicating the behavior of conventional non-smooth \acp{rl} with high fidelity. In addition, it can be duly modified to enhance the system’s dynamic control performance.
The behavior of the proposed model is demonstrated through illustrative examples as well as through a simulation of the New York/New England 16-machine 68-bus system.
\end{abstract}

\begin{IEEEkeywords}
  Rate limiter, modeling, stability analysis, nonlinear control.
\end{IEEEkeywords}

\IEEEpeerreviewmaketitle{}

\section{Introduction}

\acfp{rl} are employed in power system control loops to ensure that the rate of change of electrical and mechanical quantities remains within certain bounds, thereby contributing to system security and operational safety.  A relevant example is turbine governors, where \acp{rl} are often used to prevent the occurrence of abrupt torque changes \cite{Limitations}. 

\acp{rl} introduce discontinuities that cannot be handled within the linearization framework used to derive state-space models for power system small-signal stability analysis \cite{RL1, RL_stab}.  Consequently, their impact on the system's dynamic response is investigated exclusively using time-domain simulations. This approach requires evaluating a wide range of disturbances and scenarios, while it does not provide explicit measures of system properties, such as eigenvalues and stability margin.

In general, a \ac{rl} can be mathematically expressed through a differential equation with discontinuous right-hand, as follows:
\begin{align}
    \dot y = 
    \left\{
    \begin{aligned}
        &\min\left\{\dot y_{\max},\dot u \right\} 
        \hspace{1mm} , \ \ 
        \text{if} \
        \dot u \geq 0 \, ,  \\
        &\max\left\{\dot y_{\min},\dot u \right\} \hspace{1mm} , \ \ 
        \text{if} \ \dot u < 0 \, , 
    \end{aligned}
    \right.
\label{eq:rl}
\end{align}
where $u \equiv u(t)$ is the input signal, and $y \equiv y(t)$ the output signal of the \ac{rl}; $\dot y_{\max}>0$ and $\dot y_{\min}<0$ are, respectively, the maximum and minimum permitted rates of change for the output signal $y(t)$.  

The modeling and stability of systems with inclusion of \acp{rl} has received little attention in the power system literature.  Most existing studies are in the field of control theory, for instance, we cite \cite{RL_Main}, \cite{RL_Stab_fun}.  Despite existing studies, to the best of our knowledge, an implementation of \acp{rl} that accurately approximates \eqref{eq:rl} while being suitable for small-signal stability analysis has yet to be developed. The aim of this letter is to address this gap.

\section{Proposed Rate Limiter Model}
\label{sec:proposed}

The objective of this letter is to introduce a smooth implementation of the \ac{rl} that closely follows the response of \eqref{eq:rl}.  The proposed model is formulated as the following second-order system of differential equations \cite{bic}:
\begin{equation}
\label{eq:s1}
\begin{aligned}
\dot y &= x \, , 
 \\
\dot x &= \left(
\cmax - x 
\right) \left( x - \cmin  \right) 
\left [k_{1} (u - y) - k_{2} x \right ] - k_{3} x \, , 
\end{aligned}
\end{equation}
where $x\equiv x(t)$ is an internal state variable that governs the dynamics of the output signal $y(t)$; and $k_{1}>0$, $k_{2}>0$, $k_{3}>0$ are controlled parameters.  In this formulation, the maximum and minimum rates of change $\cmax$ and $\cmin$ are enforced smoothly by the nonlinear differential equations \eqref{eq:s1}.

The continuous nature of the proposed \ac{rl} offers significant advantages relevant to power system stability analysis and control.  First, it can be linearized around system equilibria.  In particular, at an equilibrium $(y^*,x^*)$ the derivatives $\dot y $ and $\dot x$ are zero, or equivalently, $x^{*} = 0$, $y^{*} = u^{*}$.  Linearization of \eqref{eq:s1} around $(y^*,x^*) = (u^*,0)$ gives:
\begin{equation}
  \label{eq:s1:lin}
  \begin{aligned}
    \Delta \dot  y &= \Delta x \, , \\
    \Delta \dot x &= - (k_{2} c + k_3) \; \Delta x - k_{1} c \; \Delta y + k_{1} c \; \Delta u \, ,
  \end{aligned}
\end{equation}
where $c =-\cmax \cmin$.  Equations \eqref{eq:s1:lin} can be included in the system's state matrix and thus permit accounting for \ac{rl} effects during small-signal stability analysis using well-known results from linear stability theory. For example, using these equations, one can calculate the eigenvalues of the power system model associated to the dynamic behavior of \acp{rl}.  Such an analysis is not possible with model \eqref{eq:rl}.

Another important property of \eqref{eq:s1} is that the derivative $\dot y(t)$ of the output signal is guaranteed to remain bounded within $\left(\cmin, \cmax\right)$, provided that $\cmax$ and $\cmin$ are constant.  To prove this property, we consider the energy-like Lyapunov function $V = \frac{1}{2} x^2$, with time derivative
\begin{equation}
  \dot V = ( \cmax - x ) ( \cmin - x ) [ k_{1}(u - y) x - k_{2} x^2 ] - k_{3} x^2 .
  \label{eq:dlyap} 
\end{equation}
At $\cmax$ and $\cmin$, we have:
\begin{equation}
  \label{eq:dss1}
  \begin{aligned}
    &x=\cmax: 
      \quad
      \dot V =-k_{3} \cmax^{2} \, , \\
    & x = \cmin:
      \quad
      \dot V = -k_{3} \cmin^{2} \, .
  \end{aligned}
\end{equation}
From \eqref{eq:dss1}, it is apparent that $\dot V<0$ and therefore $V$ decreases with time.  Furthermore, due to \eqref{eq:s1} being Lipchitz continuous, it is concluded that if for an arbitrary time instant $t_{0}$ it holds that $V(t_{0})\in\left(\cmin,\cmax\right)$, then for $t\geq t_{0}$, $\dot y(t) \in \left(\cmin,\cmax\right)$.  Most importantly, this is satisfied independently from the values of $k_{1}$, $k_{2}$, $k_3$.  The proof is completed.

\begin{figure}[ht!]
  \begin{subfigure}{0.24\textwidth}
    \includegraphics[width=\linewidth]{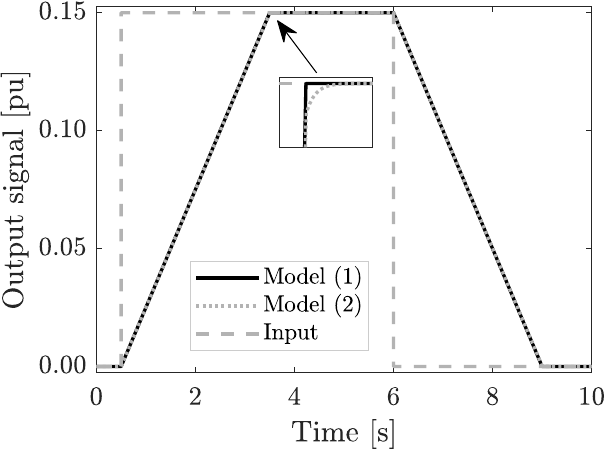}
    \vfill
    \caption{Output signal.}
    \label{fig:ex_1}
  \end{subfigure}
  \begin{subfigure}{0.24\textwidth}
    \includegraphics[width=\linewidth]{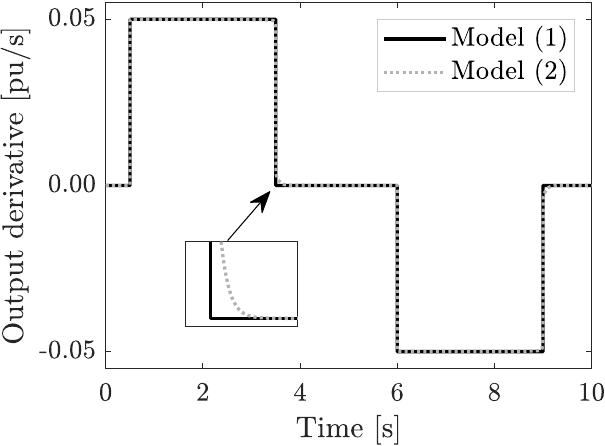}
    \centering
    \caption{Output signal derivative.} 
    \label{fig:ex_1a}
  \end{subfigure}
  \caption{Response of conventional \ac{rl} and the proposed.} 
  \label{fig:pr_conc}
\end{figure}

We illustrate the response of the proposed model to a step-change of the input signal $u(t)$ from 0 to 0.15~pu.  The upper and lower derivative limits are set to $\cmax = 0.05$~pu/s and $\cmin = -0.05$~pu/s.  
The output signal, shown in Fig.~\ref{fig:ex_1}, confirms that the proposed model adheres to the specified derivative bounds while closely following the response of model \eqref{eq:rl}, with negligible deviations.  Similar conclusions can be drawn by observing the output signal derivative shown in Fig.~\ref{fig:ex_1a}.  Moreover, as expected, the proposed \ac{rl} model leads to a smooth transition of $\dot y(t)$ from the upper limit to 0.  This is in contrast to model \eqref{eq:rl}, which imposes an abrupt derivative change from 0.05 to 0~pu/s in a single simulation step, leading to discontinuities on the first and second derivatives of the output signal.  The values of $k_{1}$, $k_{2}$, $k_3$ used in this example are $k_{1}=1800$, $k_{2}=120$ and $k_{3}=10^{-1}$.

At this point, we note the distinct role of each of 
$k_{1}$, $k_{2}$, $k_{3}$ in the transient response of the proposed model: $k_{1}$ dictates the slope of $y(t)$, with larger values leading $\dot y(t)$ to approach its limit faster following a disturbance;  $k_{2}$ dictates the damping of the model, with higher values leading to smoother transitions of $x(t)$ and consequently also of $y(t)$; finally, $k_3$ is employed to maintain $\dot y(t)$ within the desired range $\left (\cmin,\cmax\right)$.  Increasing $k_{3}$ \textit{smoothens} the transition of $\dot y(t)$ from its limits to 0.  An accurate replication of \eqref{eq:s1} typically requires $k_1 > k_2$. 
Best dynamic performance results are obtained if $k_1$, $k_2$, $k_3$ are tuned on a per-device basis.

A relevant feature of the proposed model is that it can be reconfigured so that it enhances the dynamic response of the system rather than merely replicating \eqref{eq:rl}.  This benefit arises naturally as a byproduct of the ability to adjust the response of \eqref{eq:s1} through the parameters $k_1$, $k_2$, $k_3$.  To illustrate this feature, we consider an ordinary differential equation $\dot{z}= f\left( z, u\right)$, where $z$ is the state, $u$ is controlled input, and $f$ is a nonlinear function. 
Considering $u=y$, the following modified version of the \ac{rl} model can be used to regulate a given function, $g(z,y)$, to zero, while ensuring boundedness of the input derivative: 
\begin{equation}
  \label{eq:control1}
  \begin{aligned}
    \dot y &= x \, , \\
    \dot x &= \left(
             \cmax - x \right) \left( x - \cmin  \right) 
             \left [k_{1} g\left(z,y\right) - k_{2} x \right ] - k_{3} x \, , 
  \end{aligned}
\end{equation}
Since in steady state $x^{*}=0$, we get $g\left(z^{*},y^{*}\right)=0$.  The potential of utilizing the proposed model to improve the control performance is further discussed in the next section.

\section{Case Studies}
\label{sec:case}

In this section, we study the behavior of the proposed model through numerical simulations, considering \acp{rl} for \ac{gfl} \acp{vsi}.  These case studies investigate the impact of \ac{rl} modeling on stability analysis and demonstrate the potential of using \eqref{eq:control1} to enhance the performance of converter controllers.   We also discuss the scalability of the proposed RL model through the New York/New England 68-bus benchmark system. Simulations are carried out using Matlab. 

\subsection{GFL VSI Instability Induced by Rate Limiters}

\begin{figure}[ht!]
  \begin{subfigure}{0.232\textwidth}
    \includegraphics[width=\linewidth]{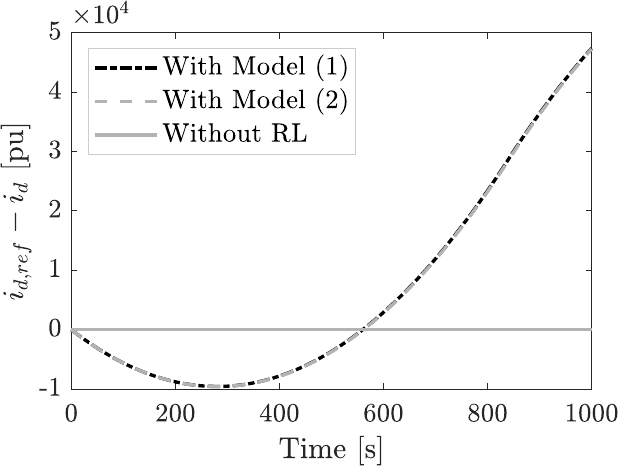}
    \vfill
    \caption{$d$-axis current error response.}
    \label{fig:norl}
  \end{subfigure}
  \begin{subfigure}{0.24\textwidth}
    \includegraphics[width=\linewidth]{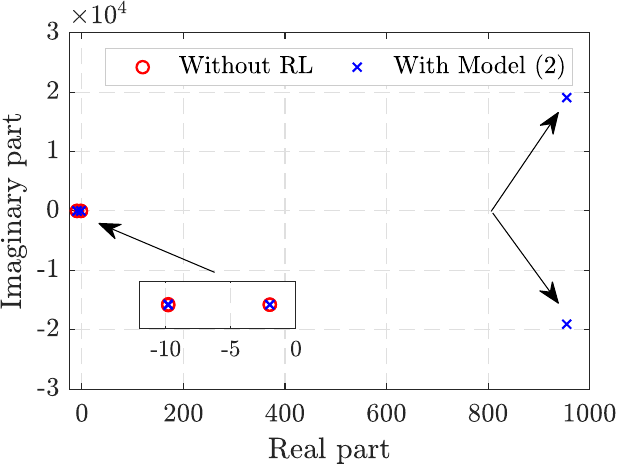}
    \centering
    \caption{Linearized system poles.} 
    \label{fig:wrl}
  \end{subfigure}
  \caption{\ac{gfl} \ac{vsi} response.} 
  \label{fig:fos}
  \vspace{-3mm}
\end{figure}

Consider a \ac{gfl} \ac{vsi} connected to an infinite bus, operating in PQ mode. The \ac{vsi} employs the control scheme presented in \cite{GFL_VSI_ex1} where inner current loops are driven by outer-layer power controllers. Without a \ac{rl}, the system remains stable following a step change in the reference active power to $10^{-3}$~pu, as shown in Fig.~\ref{fig:fos}.

Equipping the current controllers with \acp{rl} destabilizes the system.  Figure \ref{fig:norl} shows the results of time-domain simulations obtained with the conventional and the proposed RL model for $\cmax=5$~pu/s and $\cmin=-5$~pu/s.  In this case, $k_{1}=14.5\times10^{6}$, $k_{2}=2.69\times10^{3}$ and $k_{3}=10^{-1}$ mimic closely the dynamic response of \eqref{eq:rl}.    The \ac{rl}-induced instability is accurately captured with both models. 

Figure~\ref{fig:wrl} shows the eigenvalue analysis of the system with inclusion of the proposed \ac{rl} model.   If the \ac{rl} is modeled through \eqref{eq:rl}, which neglects the rate limits at the equilibrium, eigenvalue analysis is inconsistent as it leads to conclude that the system is small-signal stable.  On the other hand, the proposed RL model is able to capture the instability. 

\subsection{\ac{gfl} VSI Current Regulator}

In this section, we consider a \ac{gfl} \ac{vsi} model. The \ac{vsi} current $\ii$ is represented in the $dq$ reference frame and through feedback linearization, it can be simplified into following differential equation \cite{GFL_VSI_ex1}:
\begin{equation}
  l \dot{\ii} = -r\ii + v \, ,
  \label{ce}
\end{equation}
where $l$, $r$ are the \ac{vsi} output inductance and parasitic resistance respectively; and $v$ is a controlled voltage. 

\begin{figure}[ht!]
  \centering
  \includegraphics[width=1\linewidth]{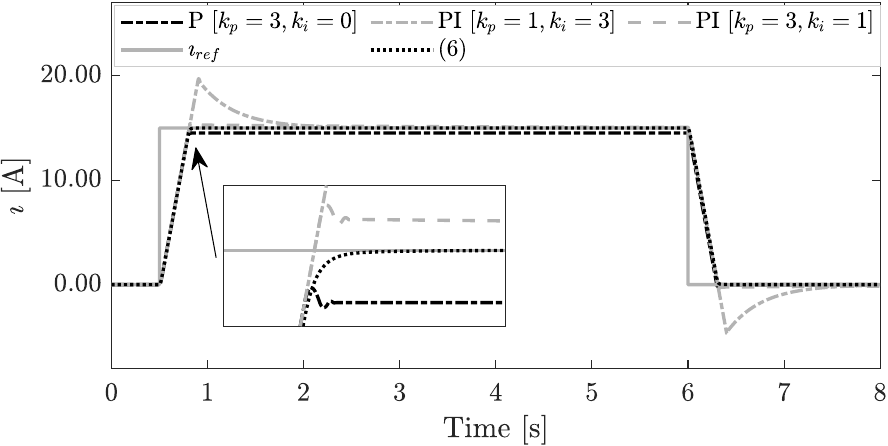}
  \vfill
  \caption{\ac{vsi} current regulated by rate-limited controller.}
  \label{fig:ex_2}
\end{figure}

Aiming to regulate the current to a reference value $\ii_{\rm ref}$ we employ a PI controller. The PI output derivative is limited through a \ac{rl} modeled through \eqref{eq:rl}. The time-domain simulation results for different values of the PI control parameters $k_p$ and $k_i$ are shown in Fig.~\ref{fig:ex_2}, where we have set $l=1$~mH, $r=0.1$~$\Omega$ and derivative bounds $\pm5$~pu/s.  The considered disturbance is a step-change of $\ii_{\rm ref}$ from 0 to 15~A.  The results indicate that the presence of the \ac{rl} induces a windup-like effect to the PI integral state, leading to overshoots/undershoots as well as to a large settling time. Moreover, increasing $k_p$ can help reduce these oscillations, yet by further increasing the settling time. A fast response can be achieved for $k_{i}=0$, however this comes with a significant steady-state error. 

We focus on showcasing an additional feature of the proposed formulation, that is, it can be used to enhance control performance. To this aim, the PI and conventional \ac{rl} limiter are substituted by a regulator in the form of \eqref{eq:control1}, wherein, $u=v$, $z=\ii$, and $g\left(\ii\right)= \ii_{\rm ref} - \ii$.  The controlled parameters are set to $k_{1}=155.5$, $k_{2}=63$ and $k_{3}=10$, while the derivative bounds are set again at $\pm5$~pu/s.  The results are shown in Fig.~\ref{fig:ex_2} and indicate that the proposed design outperforms the rate-limited PI, as it effectively regulates the current to its reference with a short settling time and without inducing overshoots.

\subsection{New England/New York 68-Bus System}

In this final example, we investigate the impact of \acp{rl} on the stability analysis of the New England/New York 16-machine 68-bus system \cite{Mat_sim}. Each machine is equipped with automatic voltage regulation, a power system stabilizer and a turbine governor. Moreover, for the needs of this study, the output torque of each turbine governor is assumed to include a \ac{rl}.

\begin{figure}[ht!]
  \begin{subfigure}{0.24\textwidth}
    \includegraphics[width=\linewidth]{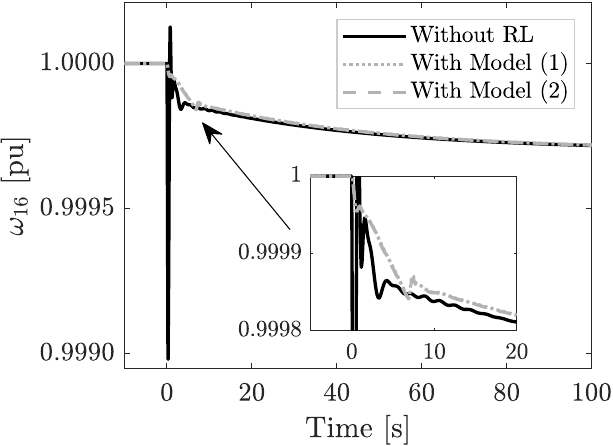}
    \vfill
    \caption{Rotor speed, 16th machine.} 
    \label{fig:speed}
  \end{subfigure}
  \begin{subfigure}{0.24\textwidth}
    \includegraphics[width=\linewidth]{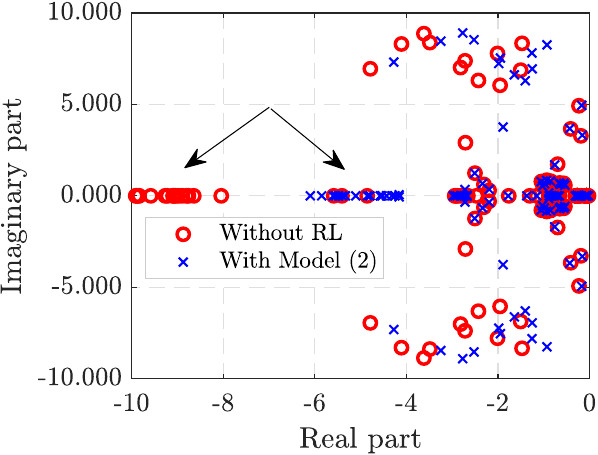}
    \centering
    \caption{Linearized system poles.} 
    \label{fig:poles}
  \end{subfigure}
  \caption{Multi-machine power system response.} 
  \label{fig:ny}
\end{figure}

We conduct a time-domain simulation of the system considering, at $t=0$~s, a $0.7$~pu increase in the power order of the 16th machine turbine governor.  Figure~\ref{fig:speed} shows the response of the 16th machine rotor speed ($\omega_{16}$), comparing three cases: (i) without \acp{rl}; (ii) with \acp{rl} modeled through \eqref{eq:rl}; and (iii) with \acp{rl} modeled through the proposed formulation \eqref{eq:s1}. \acp{rl} limits are set to $\pm0.1$~pu/s.  In the absence of \acp{rl}, the rotor speed shows significant overshoots and undershoots due to abrupt torque changes following the disturbance.  These oscillations are eliminated when \acp{rl} are included in the control loop.  Furthermore, the proposed \ac{rl} model closely replicates the behavior of \eqref{eq:rl}. 

Each \ac{rl} modeled by \eqref{eq:s1} introduces two state variables.  In this case, the system model includes 257 and 225 states, with and without, respectively, the proposed \ac{rl} model.  The increase in the computational burden is minimal, with simulation times remaining virtually unchanged.

Finally, we linearize the system around the pre-disturbance equilibrium and show its eigenvalues in Fig.~\ref{fig:poles}. As already discussed, the classical model \eqref{eq:rl} coincides with the no-limiter case in eigenvalue analysis, as its discontinuous nature prevents its influence from being incorporated into the state matrix.  In contrast, the proposed smooth model is able to capture the impact of \acp{rl} on the system's dynamic behavior, as reflected in the shifts of the associated eigenvalues. 

\section{Conclusion}
\label{sec:conclusion}

This letter introduces a \ac{rl} model suitable for power system stability analysis and control. The proposed model is continuous and smooth, which enables accounting for its impact on system dynamics when conducting small-signal stability analysis.  Moreover, its rate-limiting property is exact, in the sense that the derivative is guaranteed to remain within specified bounds regardless of the chosen controlled parameter values.  The proposed model can be suitably implemented in power system control loops to improve the accuracy of stability analysis or enhance dynamic control performance.


\end{document}